\documentclass[aps,pre,preprint]{revtex4-1}

\usepackage{graphicx}
\usepackage{amsmath,amssymb}
\usepackage{psfrag}
\usepackage[utf8]{inputenc} 
\usepackage{color}

\newcommand{\rhoa}{v_{\scriptscriptstyle \rm a}}
\newcommand{\rhoc}{v_{\scriptscriptstyle \rm c}}
\newcommand{\thetac}{\theta_{\scriptscriptstyle \rm c}}
\newcommand{\be}{\begin{equation}}
\newcommand{\ee}{\end{equation}}
\newcommand{\bea}{\begin{eqnarray}}
\newcommand{\eea}{\end{eqnarray}}

\newcommand{\lav}{\left\langle}
\newcommand{\rav}{\right\rangle}

\newcommand{\nv}{{\bf n}}
\newcommand{\tv}{{\boldsymbol \tau}}
\newcommand{\deriv}[1]{{\partial\over\partial{#1}}}

\def\e0{{\epsilon_0}}
\def\kB{{k_{\rm B}}}

\begin{document}
 
\title{Nonequilibrium Phase Transition in Constrained Adsorption}

\author{Mauro Sellitto}

\email{mauro.sellitto@unicampania.it, mauro.sellitto@gmail.com}

\affiliation{ Dipartimento di Ingegneria, Universit\`a degli Studi
  della Campania ``Luigi Vanvitelli'', Via Roma 29, 81031 Aversa,
  Italy. \\ The Abdus Salam International Centre for Theoretical
  Physics, Strada Costiera~11, 34151 Trieste, Italy.}
 \begin{abstract} 

We study the adsorption-desorption of fluid molecules on a solid
substrate by introducing a schematic model in which the
adsorption/desorption transition probabilities are given by {\it
  irreversible} kinetic constraints with a tunable violation of local
detailed balance condition.  Numerical simulations show that in one
spatial dimension the model undergoes a continuous nonequilibrium
phase transition whose location depends on the irreversibility
strength.  We show that the hierarchy of equations obeyed by
multi-point correlation functions can be closed to the second order by
means of a simple decoupling approximation, and that the approximated
solution for the steady state yields a very good description of the
overall phase diagram.
\end{abstract}

\maketitle

There is a growing interest in statistical mechanics models whose time
evolution is ruled by a Master equation with kinetic constraints
(suppressing transitions between some configuration pairs), as their
average macroscopic properties closely resembles that of more
realistic systems governed by Newton or Langevin equation of motion.
In the past decades, the study of such systems has been generally
devoted to cases in which transition probabilities satisfy global or
{\em local} detailed balance condition including aging glasses, tapped
granular materials, driven diffusion
etc.~\cite{FrAn,FrBr,RiSo,KCM_rev2}.  In the present contribution I
will consider kinetically constrained dynamics with a broken
time-reversal symmetry, namely, situations in which detailed balance
is violated at the level of single transition probability.  I show
that this new ingredient brings about nonequilibrium critical
properties even in one spatial dimension~\cite{MaDi}.  To fix ideas I
will consider the specific example of adsorption-desorption
dynamics~\cite{Evans,Talbot_rev}. This is characterized by two main
features: (1) absence of microscopic reversibility and (2) hindered
adsorption by blockage of previously adsorbed molecules. These two
crucial features are well accounted for by random sequential
adsorption models and have been thoroughly
investigated~\cite{Evans,Talbot_rev}.  To appreciate better the
differences with the present approach notice that in random parking
lot~\cite{Jin,Ben-Naim}, for example, hard rods can be placed on a
line at randomly selected position only if they do not overlap with
previously adsorbed rods, and are removed regardless their local
environment. Here, instead, adsorption is restricted by kinetic
constraints rather than steric repulsion. This means that full
coverage is achievable by a suitable sequence of adsorption-desorption
events. Moreover, desorption is also restricted but only {\em
  partially}.

Consider a fluid in contact with a thermal bath at temperature $T$ and
a particle reservoir at chemical potential $\mu$.  The fluid interact
with a solid substrate onto which fluid molecules they can be
adsorbed.  The solid substrate is represented as a lattice which can
only accomodate one molecule per site. The energy of adsorbed
molecules is $H = - \e0 \sum_i \tau_i$ with $\tau_i=0,1$ denoting the
occupation variable of the lattice site $i$.  The
adsorption-desorption dynamics is subject to irreversible kinetic
constraints embodied in the transition probabilities:
\be 
w(\tau_i \to 1-\tau_i) = \lambda \, (1-\tau_i) \, f_i + 
\tau_i \, g_i \,,
\ee
where $\lambda = {\rm e}^{(\mu-\e0)/\kB T}$ and $f_i$ and $g_i$ are
functions that restrict the adsorption and desorption at site $i$,
respectively, depending on previously adsorbed molecules on
neighbouring sites. There is no bulk diffusion.  In the following we
shall focus on one-dimensional substrate of size $N$ and consider the
specific case:
\bea
\left\{
  \begin{array}{ll}
\label{eq:constraints}
f_i = 1 - \tau_{i-1} \, \tau_{i+1} \, \\ \\
g_i = \alpha + (1-\alpha) \, f_i \,.
\end{array}
\right.
\eea
This means that adsorption at site $i$ can only occur when there is at
least one vacant neighbour, while desorption is partially restricted,
with probability $1-\alpha$, by the kinetic constraint ($\alpha \in
[0,1]$ being an irreversibility parameter that quantifies the
violation of detailed balance condition). The function $g_i$
interpolates smoothly between the equilibrium constrained dynamics
($\alpha=0$) in which adsorption and desorption transition
probabilities satisfy detailed balance ($f_i=g_i$), and the
irreversible dynamics with maximum violation of detailed balance
($\alpha=1$), in which adsorption in constrained and desorption always
occurs regardless the constraint is met or not.

We assume that the system probability distribution $p(\tv, t)$ if
governed by the master equation
\be
\deriv{t} p(\tv, t) = \sum_i \left[ w( F_i \tv \to \tv)
  \, p(F_i \tv, t) - w(\tv \to F_i \tv) \, p(\tv, t)
  \right]
\ee
where $F_i\tv$ is the configuration $\tv$ with $\tau_i$ flipped to
$1-\tau_i$, that is: $F_i \tv= (\tau_1, \cdots, 1-\tau_i, \cdots,
\tau_N)$.  One can easily deduce (see, e.g., Ref.~\cite{RiSo}) that
the average of a general observable $\phi(\tv)$ evolves in time
according to
\be
\deriv{t}\lav \phi(\tv)\rav = \sum_i \lav w(\tau_i \to 1-\tau_i) \left[
  \phi(F_i\tv)-\phi(\tv) \right] \rav .
\ee
For solvability purposes, it is useful to recast the dynamics in terms
of vacancies, so we turn to the representation in which every $\tau_i$
corresponds to the variable $1-n_i$ and dynamical evolution of the
substrate is governed by the transition probabilities:
\be 
w(n_i \to 1-n_i) = 
(1-n_i) \, \left[ \alpha + (1-\alpha) \, f_i \right] + \lambda \, n_i
\, f_i \,,
\label{eq:w}
\ee
where the constraint $f_i$ is now expressed in terms of the new
variables as $f_i = n_{i-1} + n_{i+1} - n_{i-1} \, n_{i+1}$.
To characterise the substrate phase we use the fraction of vacancies
$v$ (which is simply related to the substrate coverage $\theta$ as
$\rho=1-\theta$), and the fraction of vacancies effectively available
to adsorption $\rhoa$ (which will play the role of order parameter):
\be v = \frac{1}{N} \sum_{i=1}^N n_i \, , \qquad \rhoa =
\frac{1}{N} \sum_{i=1}^N f_i \, n_i \,. \ee
When $\rhoa = 0$ there is no further available volume to adsorption
cause of the hindering action exerted by kinetic constraints. It
should be emphasized, however, that since particles can be always
desorbed with finite probability (and subsequently adsorbed), the {\em
  inactive} phase $\rhoa = 0$ is realised by a multiplicity of
microscopic configurations in which the substrate dynamics does never
get permanently stuck (at finite temperature).
In order to determine $\rhoa$ as a function of thermodynamic variables
and the irreversibility parameter $\alpha$ we consider,
following~\cite{FoRi}, the hierarchy of $k$-point correlation
functions:
\bea
D_k = \frac{1}{N} \sum_{\ell} \lav n_{\ell} \cdots n_{\ell+k} \rav \,.
\eea
By using the obvious identities: $ (1-n_i)(1-2n_i) = 1-n_i$ and $ n_i
(1-2n_i) = -n_i$, one can easily show that to the lowest order the
correlation functions $D_k$ with $k=0,\, 1, \, 2$, obey the relations:
\bea 
\left\{
  \begin{array}{ll}
D_0 (F_i\nv)- D_0(\nv) = (1- 2 n_i)/N \\ \\
     D_1 (F_i\nv)-D_1(\nv) = (1- 2 n_i) (n_{i-1} + n_{i+1})/N \\ \\
     D_2 (F_i\nv)-D_2(\nv) = (1- 2 n_i) (n_{i-2}n_{i-1} +
n_{i-1} n_{i+1}+ n_{i+1} n_{i+2} )/N 
\end{array}
\right.
\eea
which lead to the following set of dynamical equations:
\bea 
\left\{
  \begin{array}{ll}
\dot{D}_0  =  \alpha (1-D_0) + (1-\alpha)(2 D_0 - S_1)- (1- \alpha +
\lambda) (2 D_1 - D_2) \,, \\  \\
\dot{D}_1  =  2 \alpha (D_0 - D_1) + 2 (1-\alpha) D_0 - 2
(1-\alpha+\lambda) D_1 \,, \\ \\
\dot{D}_2  =  \alpha ( 2 D_1 - 3 D_2 + S_1) + (1-\alpha)( 2 D_1 +
S_1) - 3 (1-\alpha+\lambda) D_2 \,,
\end{array}
\right.
\label{eq:kinetics}
\eea
where 
\bea S_1 = \frac{1}{N} \sum_{\ell} \lav n_{\ell-1} n_{\ell+1} \rav \,.
\eea
We are interested here to the system steady state, which is obtained
by setting the above time derivatives to zero. The related system of
algebraic equations can be closed once a suitable approximation for
the two-point correlation $S_1$ is made.  The simplest one is a
decoupling approximation which neglects correlation between
next-nearest neighbouring sites and replace the average of a product
with a product of averages, $\lav n_{\ell-1} n_{\ell+1} \rav \simeq
\lav n_{\ell-1} \rav \lav n_{\ell+1} \rav $, giving $S_1 \simeq
D_0^2$. Plugging this approximation in the above equations one gets:
\bea 
\left\{
  \begin{array}{ll}
0  =  \alpha (1-D_0) + (1-\alpha)(2 D_0 - D_0^2)- (1- \alpha +
\lambda) (2 D_1 - D_2) \,, \\ \\ 0  =  2 \alpha (D_0 - D_1) + 2
(1-\alpha) D_0 - 2 (1-\alpha+\lambda) D_1 \,,\\ \\ 0  =  \alpha ( 2
D_1 - 3 D_2 + D_0^2) + (1-\alpha)( 2 D_1 + D_0^2) - 3 (1-\alpha+\lambda)
D_2 \,.
\end{array}
\right.
\eea
By writing $v=D_0$ and $\rhoa=2D_1-D_2$, and eliminating the
dependence on $\lambda$ in favour of that on $\alpha$ and $v$, one
finally finds that $\rhoa$ satisfies the quadratic equation:
\be 
\label{eq:quadratic}
A \rhoa^2 + B \rhoa + C = 0 \, 
\ee
with coefficients $A,\,B,\,C$ depending on $v$ and $\alpha$ as
follows:
\bea 
\left\{
  \begin{array}{ll}
A = 3 \alpha^2 , \\ \\ B = 2 v + \alpha \left[ (6 \alpha-5)
  v^2 + 6(1-3 \alpha) v + 6\alpha \right] , \\ \\ C = \left[
  (1-\alpha) v^2 +(3 \alpha-2) v- \alpha \right] \left[ (2-3
  \alpha) v^2 + 9 \alpha v - 3 \alpha \right] .  
\end{array}
\right.
\eea 
To locate the critical line, $\rhoc(\alpha)$, one has to set $\rhoa=0$
in Eq.~\eqref{eq:quadratic}. This amounts to solve the quartic
equation $C=0$, which has a unique physical solution:
\bea \rhoc(\alpha) &=& \frac{9 \alpha - \sqrt{45\alpha^2 + 24
    \alpha}}{6\alpha-4} .
\label{eq:rhoa}
\eea 
In the equilibrium limit $\alpha=0$ one gets $\rhoc=0$ and there is no
critical behavior as expected for a one-dimensional system in thermal
equilibrium. As soon as the irreversibility parameter $\alpha$ is
nonzero a continuous nonequilibrium phase transition sets in.  Its
location, defined by the line $\rhoc(\alpha)$, behaves as
$\sqrt{\alpha}$ for small $\alpha$ and increases monotonically.  The
latter feature can be intuitively understood by observing that the
stronger the violation of detailed balance the faster the rate at
which irreversible desorption events occur. These events will
typically destroy the previously realised local correlation of
particle arrangements on the substrate, eventually leading to a
sub-optimal coverage with respect to the equilibrium case (which is
unity, that is $\rhoc=0$). In the full irreversible case, $\alpha=1$,
one has $ \rhoc= \frac{9 - \sqrt{69}}{2} \simeq 0.346688...$ ($\thetac
\simeq 0.653312$). The Taylor expansion of $\rhoa$ near the critical
vacancy density $\rhoc$ shows that the order parameter increases
linearly in both $v-\rhoc$ and $\alpha$, so the order parameter
critical exponent is $\beta=1$, and the phase transition does arguably
belong to the universality class of {\it absorbing} phase
transition~\cite{MaDi}. Note, that the inactive (or absorbing) phase
comprises an exponential (in the system size) multiplicity of
microscopic configurations (or absorbing states) whose entropy can be
easily computed (see Eq.~(11) in Ref.~\cite{CrRiRoSe}).
The way in which the irreversible dynamics samples such configurations
near the critical threshold is a relevant problem for the Edwards
ergodic hypothesis of granular matter~\cite{Makse_rev} which, for
one-dimensional systems related to the present one, has been addressed
in Refs.~\cite{Brey,BPS,PB,Dean,Berg,Smedt}.  Although very
interesting this issue will not be considered here.

\begin{figure}[htbp]
  \includegraphics[scale=.6]{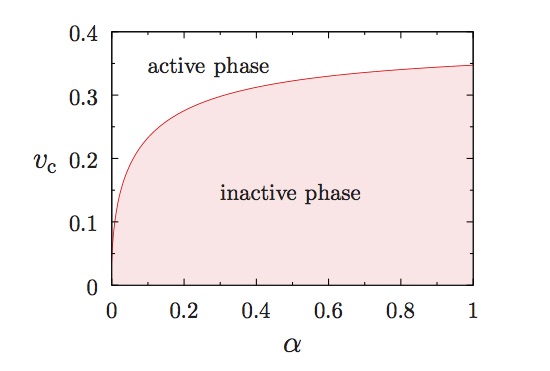}
  \caption{Nonequilibrium phase diagram of constrained adsorbtion
    dynamics: critical vacancy fraction $\rhoc=1-\thetac$
    vs. irreversibility parameter $\alpha$.}
  \label{Fig:rhoc}
\end{figure}

\begin{figure}[htbp]
  \includegraphics[scale=.6]{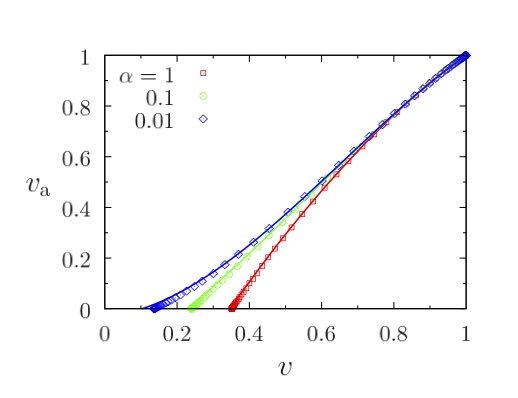}
  \caption{Fraction of available vacancies $\rhoa$ vs. vacancy
    fraction $v$ for some values of the irreversibility parameter
    $\alpha$ (full lines are theoretical results, symbols are Monte
    Carlo data).}
  \label{Fig:rhoa}
\end{figure}

\begin{figure}[htbp]
  \includegraphics[scale=.6]{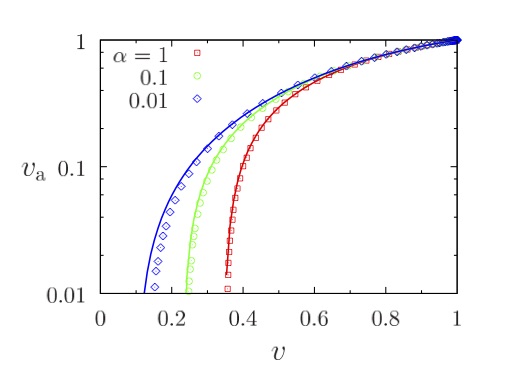}
  \caption{Same data as in Fig.~\ref{Fig:rhoa} in a log-scale plot
    (highlighting the discrepancy between theoretical and numerical
    results at small $\rhoa$).}
  \label{Fig:rhoa_log}
\end{figure}

The full phase diagram obtained in the above approximation is shown in
Fig.~\ref{Fig:rhoc}.
To assess the limit of our approximation we perform standard
grand-canonical Monte Carlo simulations for a $1D$ substrate of size
$N=2^{17}$ by using annealing rate sufficiently low as to avoid
hysteresis effect in cooling-heating cycles (typically
$10^4-10^5$ MC sweeps per unit of chemical potential).
In Fig.~\ref{Fig:rhoa} we compare the virtually exact MC simulations
of $\rhoa$ vs $v$, for some values of the irreversibility parameter
$\alpha$, with the solution of Eq.~\eqref{eq:quadratic}. The agreement
turns out to be excellent. A closer look at the data shows that
discrepancies (necessarily expected) near the threshold can be
appreciated in a log-scale plot when the irreversibility parameter
$\alpha$ becomes vanishingly small, see Fig.~\ref{Fig:rhoa_log}, i.e.,
when the equilibrium limit of constrained adsorption-desorption
dynamics is approached. Thus, the decoupling approximation becomes
increasingly accurate in the limit of fully unconstrained
desorption. Next-nearest neighbours correlation are, however, very
important in determining the kinetic approach to the steady state, as
the decoupling approximation does not close the dynamical
hierarchy~\eqref{eq:kinetics}. The problem of addressing kinetics with
a clever approximation scheme is therefore left to future work.

To summarise we have introduced a simple model of
adsorption-desorption dynamics with irreversible kinetic constraints
and studied its steady state in one spatial dimension. We found a
continuous nonequilibrium phase transition that is nicely captured by
a naive mean-field approximation.  In higher spatial dimensions a
similar nonequilibrium transition should arguably occur, and it would
be particularly interesting to investigate {\it cooperative} models in
which adsorption is promoted by two or more nearby vacancies. We
expect, in this case, a more subtle influence of the broken
time-reversal symmetry on dynamics which, on a Bethe lattice, should
lead to a nontrivial competition between hybrid and absorbing phase
transitions.  Including polydispersity in the present
  approach is also possible~\cite{Prados_mixture}, and could give rise
  to rather complex phase behaviours~\cite{ArSe}.

\end{document}